\documentclass{raa01}            

\usepackage{natbib}
\bibpunct{(}{)}{;}{a}{}{,}
\usepackage{graphicx,times}             

\newcommand{\Mpc}{\ensuremath{{\rm Mpc}}}

\begin{document}

   \title{Cosmological constraint on Brans-Dicke Model
}

   \volnopage{{\bf 2015} Vol.\ {\bf 15} No. {\bf 12}, 2151--2163~
 {\small doi: 10.1088/1674--4527/15/12/003}}
      \setcounter{page}{2151}

   \author{Ji-Xia Li
      \inst{1,2}
     \and Feng-Quan Wu
      \inst{1}
     \and Yi-Chao Li
      \inst{1}
     \and Yan Gong
      \inst{1}
   \and Xue-Lei Chen
      \inst{1,3}
   }

   \institute{National Astronomical Observatories, Chinese Academy of Sciences,
             Beijing 100012, China; {\it xuelei@cosmology.bao.ac.cn}\\
        \and
             University of Chinese Academy of Sciences,
             Beijing 100049, China
        \and
             Center of High Energy Physics, Peking University,
             Beijing 100871, China\\
\vs\no
   {\small Received 2015 April 25; accepted 2015 April 29}}

\abstract{  We {combine} new Cosmic Microwave Background (CMB){
data} from Planck {with} Baryon Acoustic Oscillation (BAO) data to
constrain the Brans-Dicke (BD) theory, in which the gravitational
constant $G$ evolves with time. {Observations of} type Ia
supernova{e} (SN{e}Ia) provide another important set of cosmological
data, as {they} may be regarded as standard candle{s} after some
empirical corrections. However, in {theories that include} modified
gravity {like} the BD theory, there is some risk and complication
{when} us{ing} the SNIa data because {their} luminosity may depend
on $G$.  In this paper{,} we assume a power law relation between the
SNIa luminosity and $G$,  but treat the power index as a free
parameter. We then test whether the difference in distances measured
with SNIa{ data} and BAO data can be reduced in such {a }model. We
also constrain the {BD} theory and cosmological parameters by making
a global fit with the CMB, BAO and SNIa data set. For the
CMB+BAO+SNIa data set, we {find} $0.08\times10^{-2} < \zeta
<0.33\times10^{-2} $ at{ the} 68\% {confidence level (}CL{)} and
$-0.01\times10^{-2} <\zeta <0.43\times10^{-2} $ at{ the} 95\%\ {CL},
where $\zeta$ is related to the {BD} parameter $\omega$ by
$\zeta=\ln(1+1/\omega)$. \keywords{cosmology: supernovae
--- Brans-Dicke --- standard-candle --- BAO} }

\authorrunning{J.-X. Li et al.}
\titlerunning{Cosmological Constraint on Brans-Dicke Model}

   \maketitle
\section{Introduction}           
\label{sect:intro}

Einstein's theory of general relativity (GR) is a major pillar of
modern physics and astronomy. As such, it is very important to
rigorous{ly} test this theory.  {Considering} the{ problems posed
by} dark matter and dark energy, i.e. on the{ scale of the} galaxy
and {larger}, gravity is dominated by unknown components{. I}t is
especially important to compare it with competing models, namely
{theories that include} modified gravity. The Brans-Dicke (BD)
theory \citep{Brans:1961sx} is the simplest{ example} of such{ a}
theory. In the BD theory, the gravitational constant $G$ is no
longer a constant, but{ rather} a scalar field which varies over
space and time. The action of the BD theory is given by
\begin{equation}
\label{action} {\mathcal S}=\frac{1}{16\pi} \int d^4 x
\sqrt{-g}\left[-\phi R+
\frac{\omega}{\phi}g^{\mu\nu}\nabla_{\mu}\phi \nabla_{\nu}\phi
 \right] +{\mathcal S}^{(m)}\, ,
\end{equation}
where $\phi$ is the BD field,  $\omega$ is  a dimensionless
parameter,   and ${\mathcal S}^{(m)}$ is the action for ordinary
matter fields ${\mathcal S}^{(m)}= \int d^4 x \sqrt{-g} {\mathcal
L}^{(m)}.$ For convenience, we can also introduce a dimensionless
field $\varphi=G \phi$. To be consistent with lab experiments, the
value of $\varphi$ at {the }present day should be $\varphi_0 =
\frac{2\omega + 4}{2\omega + 3}$ where $\omega$ is a dimensionless
parameter. In the limit{s} $\omega \to \infty, ~\dot{\varphi} \to
0${ and}$ ~\ddot{\varphi} \to 0$, BD theory {asymptotically
approaches} GR theory.

With the advent of precis{e} cosmological observations,
{c}osmological observations such as the Cosmic Microwave Background
(CMB) can be used to test the BD theory \citep{Chen:1999qh}. We have
derived limits on the BD parameter using{ data from} the Wilkinson
Microwave Anisotropy Probe (WMAP) \citep{2010PhRvD..82h3002W} and
Planck \citep{2013PhRvD..88h4053L}. Similar studies {have} also{
been} carried{ out} by others
\citep{2011IJMPS...1..183N,2005PhRvD..71j4025A,2014PhRvL.113a1101A}.
So far{,} these tests all yield results which are consistent with GR
within observational error.

{Although} the{ theories of} modified gravity affect cosmology in various ways, in many cases the
 most direct and apparent effect is on the cosmic expansion history.  The variation of $G$ over
 time induce{s} changes in the cosmic expansion rate $H(z)$ at different redshifts,
 \begin{equation}
      \label{modified FLRW}H^2 = \frac{\kappa }{3\varphi} \rho
         +\frac{\omega}{6}\left(\frac{\dot{\varphi}}{\varphi}\right)^2
            -H \frac{\dot{\varphi}}{\varphi} \, .
             \end{equation}

 Both the type Ia
supernova{e} (SN{e}Ia) and Baryon Acoustic Oscillation (BAO) data
provide means to measure distances on cosmological scales and
 supplement the CMB in test{s} of{ theories of} modified gravity (for a review, see e.g. \citealt{2013arXiv1309.5382K}).
  However, {although} the BAO data {represent} a geometric measurement of distance
 and can be simply applied to test {a} modified gravity
 model,  there is some risk and complication {when} applying the SNIa data in modified
 gravity models. The problem {in} the underlying principle of distance measurement {with} SN{e}Ia is that
 they can be considered standard candles after applying a correction based on the
 Phillips relation \citep{1993ApJ...413L.105P}, which links the SNIa luminosity to the time scale of the
 light curve. However, this is an empirical fact derived from nearby
 {SNe}. It is generally argued that the reason {for} this is that the critical mass of the
 accreting white dwarf is close to the Chandrasekhar mass, but a supernova explosion is
 a highly complicated and nonlinear process{ in which} the composition, spin and accretion of the white dwarf
 all var{y}. So far{,} the explosion process {has} not{ been} fully understood \citep{2010NewAR..54..201H}.
 In {fact}, even the nature of the progenitor of {an} SNIa {is} still hotly
 debated \citep{2012NewAR..56..122W,2014ARA&A..52..107M}.
 It is quite conceivable that the variation of $G$ may affect the{ light observed from an} {SN} explosion in unknown ways,
 and thus cause{ a} systematic deviation from the local Phillips relation and bias the measurement.
 Indeed, the question of whether there is a systematic difference between distances measured with
 SN{e}Ia and BAO {has been} investigated by a number of authors, and recent studies generally show that the
 two data sets are consistent with each other  \citep{2009JCAP...06..012A,2011JCAP...09..003E,2012PhRvD..85b3517W,2014PhRvD..90h3006C,2014arXiv1401.0046M}, although at the $2\sigma$ level, there might still be some {disagreement}, especially when considering the high-redshift Ly$\alpha$ BAO measurement \citep{2014arXiv1411.1074A,2015A&A...574A..59D,2015arXiv150100796N}.
Future improvements {in} measurement precision may reduce or
sharpen such {a }difference.
 Because of this concern, in our previous studies \citep{2010PhRvD..82h3002W,2013PhRvD..88h4053L},
 we have refrained from using the SNIa data.

  However, the SNIa data provide very powerful tests of cosmological models, and {they} could
 significantly improve the precision in the measurement of cosmological parameters.
 A number of SNIa data{ sets} have been collected by{ supernova surveys like} the SCP \citep{2012ApJ...746...85S}, SNLS \citep{2006A&A...447...31A}, ESSENCE \citep{2008ApJ...689..377W}, CANDELS \citep{2011ApJS..197...35G}, CLASH \citep{2011AAS...21722706P}, {and }SDSS-II \citep{2009ApJS..185...32K}. On-going surveys, such as the Nearby Supernova Factory \citep{2004NewAR..48..637W}, Palomar Transient Factory \citep{2014MNRAS.444.3258M}, La Silla/QUEST \citep{2012IAUS..285..324H},
 PanSTARRS \citep{2014ApJ...795...45S} and DES \citep{2013APh....42...52G}, {will} also contribute {additional} SNIa data. In the future, LSST \citep{Abell:2009aa} will greatly increase the number of{ identified} SN{e}Ia. It is therefore interesting and important to consider{ also} applying {these data} in the test of
 modified gravity models.

 The precise form of how SNIa luminosity varies
 with $G$ is unknown. Assuming that SNIa luminosity is proportional to the Chandrasekhar mass,
 $L \sim G^{-3/2}$, here we will consider more general possibilities.  If the variation of $G$ is small,
  we could parameterize the effect by assuming that after making
 the correction based on{ the} local Phillips{ }relation, the SNIa luminosity depends on
 $G$ with $L \propto G^{-\gamma_G}$, where $\gamma_G$ is a free parameter.

The expansion of the Universe in the BD  case can be obtained by
solving Equation~(\ref{modified FLRW}). The CMB angular power
spectrum can be calculated with the public Boltzmann code
\texttt{CAMB} \citep{2002PhRvD..66j3511L}, and a modified version
include{s} {an implementation} of the BD model
\citep{2010PhRvD..82h3002W}. Constraint{s} on the BD parameter and
other cosmological parameters can be obtained from a modified
version \citep{2010PhRvD..82h3003W}  of the Markov-Chain Monte Carlo
code \texttt{CosmoMC} \citep{2002PhRvD..66j3511L} which uses the
\texttt{CAMB} as {the }CMB driver.
 In this paper, we shall also include SNIa data in the tests.

The cosmological distances measured with SNIa and BAO data are most commonly used in current research.
{Although t}here are also other probes of cosmic expansion, e.g., the observational Hubble
parameter \citep{2010PhLB..689....8Z,2011ApJ...730...74M,2013PhRvD..88j3528Y},
we will limit ourselves to the SNIa and BAO data in the present  paper.

\section{The Data}
We {will} use CMB, BAO and SNIa data in this paper. For the CMB
data, we compare the angular  temperature and polarization power
spectrum predicted with our modified \texttt{CAMB} code with the
Planck 2013 data
\citep{2010PhRvD..82h3002W,2010PhRvD..82h3003W,2013PhRvD..88h4053L}.
For the distance measurements, the comoving distance to redshift $z$
{i}n the flat  Friedmann-Robinson-Walker (FRW) model is
given by
\begin{equation}
\label{eq:DCdefine} D_C(z) = \int_0^z \frac{c dz'}{H(z')}\, ,
\end{equation}
where $H(z')$ is the expansion rate at redshift $z'$.

\subsection{BAO Data}
\label{sec:BAO_data}

The BAO distance measurement is derived from observations of large
scale structure.  Acoustic oscillations before {r}ecombination left
wiggles on the correlation function and power spectrum at specific
distance scales. For a given cosmological model, such {a }distance
scale can be predicted. The galaxy correlation function and/or power
spectrum are measured within a redshift range in a galaxy survey,
and if the baryon wiggles are detected, they provide a measurement
of the corresponding distance scale. In principle, the distance
measured from the large scale structure modes perpendicular to the
line of sight provides a measurement of the angular diameter
distance $D_A(z)$, while the distance measured from modes along the
line of sight connects the redshift distance  to the physical
distance, i.e. $H(z)$ can be derived from it. In practice, however,
measurements are often made by combining all modes to suppress the
noise, and the distance derived is the volume weighted distance
$D_V(z)$, which is given by
\begin{equation}
\label{eq:DVdefine} D_V(z) = \left[\frac{cz}{H(z)} (1+z)^2
D_A(z)^2 \right]^{\frac{1}{3}} = \left[\frac{cz}{H(z)} D_C(z)^2
\right]^{\frac{1}{3}}\,,
\end{equation}
where $D_A$ is the angular diameter distance. The volume distance
is related to the comoving distance~by
\begin{equation}
\label{eq:DCfromDV} D_C(z) = \sqrt{\frac{H(z)D_V(z)^3}{cz}}\,,
\end{equation}
and the corresponding measurement error is
\begin{equation}
\label{eq:DCfromDVerr} \Delta D_C(z) = \sqrt{\frac{9H(z)
D_V(z)}{4cz} \Delta D_V^2(z)}\,.
\end{equation}
Below we summarize the BAO data used in this study.

\begin{enumerate}

\item[(1)] {The 6dF galaxy data at $z=0.106$}.  \citet{2011MNRAS.416.3017B} analy{z}ed the BAO signal with {a }large-scale correlation function of the 6dF Galaxy Survey (6dFGS). They measured the volume distance $D_V(z_\mathrm{eff}) = 457 \pm 27 ~\mathrm{Mpc}$ and the distance ratio $r_d/D_V(z_\mathrm{eff}) = 0.336 \pm 0.015$ at an effective
redshift $z_\mathrm{eff}=0.106$ where $r_d$ is the comoving sound horizon at the baryon-drag epoch.

\item[(2)] {The SDSS DR7 main galaxy sample at $z=0.15$}.  \citet{2015MNRAS.449..835R} determined the volume
distance  to be $D_V(z_\mathrm{eff} = 0.15) = (664 \pm 25)(r_d /
r_{d,{\rm fid}}) ~\mathrm{Mpc}$ with SDSS DR7.

\item[(3)] {The joint SDSS DR7 and 2dF galaxy data at $z=0.275$}. \citet{2010MNRAS.401.2148P} gave a joint analysis
 by including the SDSS DR7 galaxy sample and 2-degree Field Galaxy Redshift Survey{ }(2dFGRS) data.
 They reported the distance to be $r_d/D_V = 0.1390 \pm 0.0037$ at redshift $z=0.275$.

\item[(4)] {The SDSS DR11 galaxy data at $z=0.32$}. Using the SDSS III DR11 sample, \citet{2014MNRAS.441...24A}
measured the correlation function and power spectrum, including
density-field reconstruction of{ the} BAO feature. The best fitted
result gave $D_V(z=0.32) = (1264 \pm 25 \mathrm{Mpc})(r_d/r_{d,{\rm
fid}})$ in their fiducial cosmology{ with} $r_{d,{\rm fid}} =
149.28{\,} \mathrm{Mpc}$.

\item[(5)] {The SDSS DR7 LRG data at $z=0.35$}. \citet{2012MNRAS.427.2168M} reported a $1.9\%$
measurement of the distance ratio $D_V (z = 0.35)/r_d = 8.88 \pm
0.17$ by using a reconstruction technique on the SDSS DR7 red
luminous galaxy (LRG) dataset.

\item[(6)] {The SDSS DR9  LRG data at $z=0.57$}. \citet{2012MNRAS.427.3435A} used the SDSS III DR9 CMASS LRG
sample to reconstruct the BAO feature. They reported a distance
ratio{ of} $D_V/r_d = 13.67 \pm 0.22$ at redshift $z = 0.57$.

\item[(7)] The WiggleZ galaxy data at  $z$ = (0.44, 0.6, 0.73). The WiggleZ Dark Energy Survey data were analy{z}ed by  \citet{2014MNRAS.441.3524K}. They measured the model independent distance $D_V(r_{d,{\rm fid}}/r_d)$ = ($1716 \pm 83$\,{Mpc}, $2221 \pm  101$\,{Mpc}, $2516\pm 86$\,{Mpc}) at
effective redshifts $z$ = (0.44, 0.6, 0.73), respectively. Note that
\texttt{CosmoMC} uses acoustic parameter $A(z) \equiv
D_V(z)\sqrt{\Omega_m H_0^2}/cz$ introduced by
\citet{2005ApJ...633..560E}.

\item[(8)] {The SDSS DR11 Ly$\alpha$ data at $z=2.36$}. \citet{2014JCAP...05..027F} analy{z}ed the SDSS III DR 11 data
and studied the cross-correlation of quasars with the Ly$\alpha$
forest absorption. At redshift $z=2.36$, they reported a
measurement of the BAO scale along the line of sight $c/(H r_d) =
9.0\pm 0.3$ and across the line of sight $D_A / r_d = 10.8 \pm
0.4$. We can transform them to the volume distance ratio $D_V /
r_d = 30.35 \pm 0.822$ from Equation {(}\ref{eq:DVdefine}{)}.
\end{enumerate}

\subsection{The SNIa Data}
\label{sec:SNIa_data}

We use the updated Union2.1 compilation{ of} SN{e}Ia\footnote{Available at
\textit{http://supernova.lbl.gov/Union}}. This data set includes 580
SN{e}Ia with redshift $z$, covering a range from 0.015 to 1.414. For
each SNIa in the sample, the redshift, the distance modulus $\mu$
and its error estimate $\Delta \mu$ \citep{2012ApJ...746...85S}
{are} given. The distance modul{us} $\mu \equiv m - M$ is given by
\begin{equation}
\label{eq:mu_origin} \mu = 25 + 5\log_{10} D_L(z) + K(z) + A\,,
\end{equation}
where $m$ is the apparent magnitude at peak luminosity,  $M$ is the
absolute magnitude after{ the} correction {based on} {the }light{ }curve{ }shape
with the SALT2 model \citep{2007A&A...466...11G},{ and} $D_L(z)$ is the
luminosity distance in $\Mpc$. $K(z)$ is the K-correction{ and}  $A$ is
the extinction{;} these{ terms} are not relevant for our discussion below and
we shall neglect them.

It is argued that the peak luminosity of{ an} SNIa is determined by the Chandrasekhar mass limit, which satisfies $\mathcal{M}_\mathrm{limit} \propto G^{-3/2}$ \citep{1999astro.ph..7222A,2006IJMPD..15.1163G}. Based on the fact that $L\propto \mathcal{M}_\mathrm{limit}$, a modification should be added to the absolute magnitude of{ an} SNIa when a theory with a varying $G$ is considered. Us{ing} {the }definition of magnitude,
\begin{eqnarray}
M &&= -2.5 \log \frac{L}{4\pi D_L^2}\,.
\end{eqnarray}
If $L \propto G^{-3/2}$, we should have
\begin{eqnarray}
\label{mu_G} \mu   &&= \mu_{\rm geo} -
\frac{15}{4}\log\frac{G}{G_0}\,,
\end{eqnarray}
where $\mu_{\rm geo}$ is {a} purely geometric distance
modulus\footnote{Note that $\mu_{\rm geo}$ is not exactly the $\mu$
value for the GR case, because in the BD theory the expansion
history is also changed.}, while the second term is due to the
variation of SNIa luminosity induced by the change in $G$.

As the SNIa explosion mechanism is still not completely understood, here we consider a more general relationship
between the peak luminosity and $G$. We parameterize the relation as
$L \propto G^{-\gamma_G}$,   then
\begin{eqnarray}
\label{eq:mu_gammaG}
\mu(z) &=& \mu_{\rm geo} (z)- \frac{5}{2} \gamma_G \log \frac{G}{G_0}\,, \\
\label{eq:mu_brans}
       &= &\mu_{\rm geo}(z) + \frac{5}{2} \gamma_G \log \phi(z)\,.
\end{eqnarray}
This difference of $\Delta \mu = \frac{5}{2}\gamma_G\log\phi$ is the correction which should be added to the distance modulus, or equivalently the absolute magnitude of SNIa data in the case of{ a} varying-$G$.

The true luminosity distance should be
\begin{equation}
D_L= 10^{ [ \mu_{\rm geo} - 25 ] / 5}[\mathrm{Mpc}],
\end{equation}
and to compare with the BAO data, we can convert this to the comoving distance by
assuming a flat $\Lambda$ FRW model.
\begin{equation}
\label{eq:DCfromDL} D_C = \frac{D_L}{1+z}\,.
\end{equation}
However, if we do not know the effect of varying $G$ on SN{e}Ia, we
would get an {\it apparent luminosity distance},
\begin{equation}
D_L^{\rm app}= 10^{ [ \mu - 25 ] / 5}[\mathrm{Mpc}]\,.
\end{equation}
{S}imilarly, this can be converted to an apparent
comoving distance $D_C^{\rm app}$. There are many SNIa data points,
but for each data point the measurement error is very large. To
better visualize the data, and also to reduce the amount of
computation, we group the 580 SNe into 20 redshift bins. In the
$j$-th bin of the binned data, the mean value of the comoving
distance is given as
\begin{equation}
\label{eq:sn_bin} \bar{D}_C(\bar{z}_j) = \frac{\sum_i{w_i
(D_C)_i}}{\sum_i{w_i}}\,,
\end{equation}
where $w_i$ is the inverse of {the }comoving distance error of each
supernova, that is $w_i = 1 / \epsilon_i${,} {a}nd $\bar{z}_j$ is
the {central} value of the binned redshift. The error of the $j$-th
comoving distance is then given by \be \epsilon(z_j) = \sqrt{
\frac{\sum_i{w_i \left[ \bar{D}_C(\bar{z}_j) - (D_C)_i
\right]^2}}{\sum_i{w_i}} }\,.  \ee

The log likelihood is given by $\mathcal{L}=-\chi_{SN}^2/2$, where
\begin{equation}
\label{eq:sn_chisq} \chi_{SN}^2 = \sum_i \frac{(\mu_{\rm th}(z_i)
- (\mu_{\rm obs})_i)^2}{(\sigma_{\mu})^2_i}\,{.}
\end{equation}
$\mu_{\rm th}(z_i)$ is the distance modulus at redshift $z_i$
calculated with the theoretical model and $\mu_{\rm obs}$ is its
observed value.

\section{Results}
\label{sect:result} Historically, the BD model is parameterized with
the parameter $\omega$. However, this is inconvenient to use
{because} the GR case is {included} in the limit of $\omega \to
\infty$. In
 \citet{2010PhRvD..82h3003W}, we introduced the parameterization
\begin{equation}
\label{eq:zeta_define} \zeta=\ln\left(1+\frac{1}{\omega}\right)\,,
\end{equation}
where the GR limit is  $\zeta \to 0$. In principle, {an
}arbitrary value of $\omega$ or $\zeta$ is allowed, but the
\texttt{CAMB} code can only run effectively when $\zeta$ is
relatively small, so that the deviation from the $\Lambda$CDM model
is not too much. The fitting range of $\zeta$ is set to $(-0.014,
0.039)$, the same as \citet{2010PhRvD..82h3003W} used.

\subsection{The Redshift-{D}istance Relation}

\begin{figure}
\centering
\includegraphics[width=0.7\textwidth, angle=0]{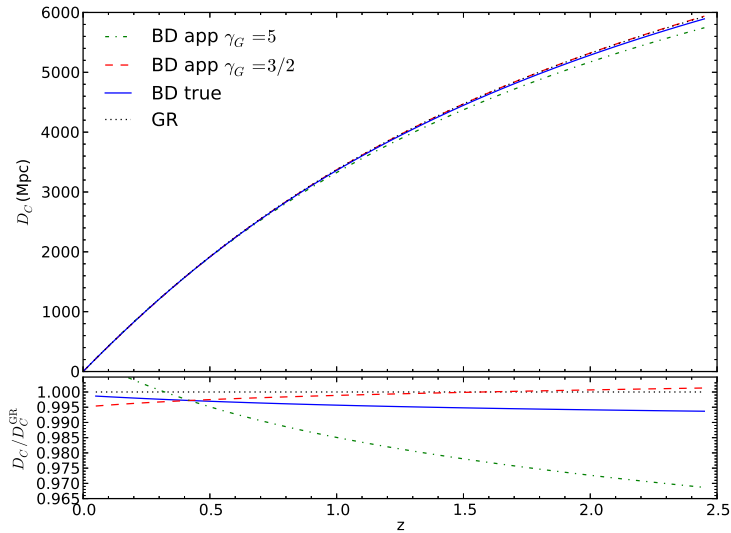}

\caption{\baselineskip 3.6mm \label{fig:dC_gammaG_effect} The
redshift - comoving distance relation for the $\Lambda$CDM model
({\it black dotted line}), {the }BD model with $\zeta=0.01$ ({\it
blue solid line}) and the apparent comoving distance  for
$\gamma_G=3/2$ ({\it red dash{ed} line}) and $\gamma_G=5$ ({\it
green dash-dot{ted} line}). }
\end{figure}

First let us consider the effect of {the }BD model on {the }comoving
distance. In Figure~\ref{fig:dC_gammaG_effect}, we plot the comoving
distance for the $\Lambda$CDM model, the (true) comoving distance
for the BD model with $\zeta=0.01$, and the apparent comoving
distance for $\gamma_G=3/2$ and $\gamma_G=5$, as well as the
relative difference with respect to the GR model. As we can see,
there are some differences in the redshift-distance relation between
the GR model and {the }BD model.  If we know how the SNIa luminosity
is affected by the variation
 of  $G$ and take this into account as in Equation~(\ref{eq:mu_brans}), using the SNIa{ data} we would have obtained the
 true comoving distance just as those derived from the BAO data. However, if this variation in SNIa luminosity is not taken into account, we would then get the apparent comoving distance, and {it }deviates more significantly from the GR redshift-distance relation.
Also, as we shall see below, the correction term $\frac{5}{2}\gamma_G \log \phi$ {requires} a large $\gamma_G$ to {have a} significant effect, due to the fact that in BD theory $\phi$ is usually within the range $1 \pm 10^{-3}$.
Here in the global fit we choose the prior to be $\gamma_G \in (-90, +90)$.

\begin{figure}
\centering
\includegraphics[width=0.65\textwidth]{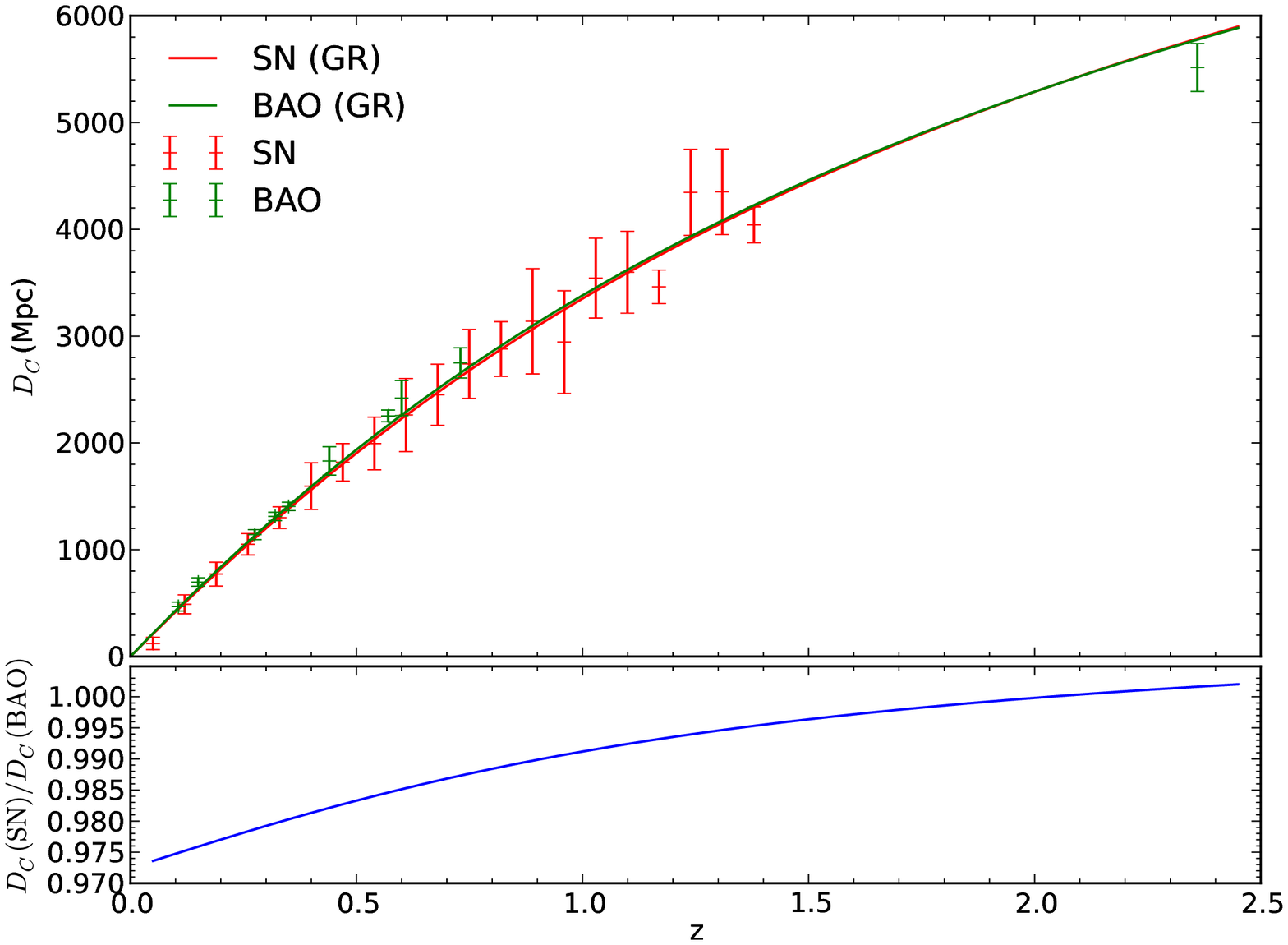}

\caption{\baselineskip 3.6mm \label{fig:dC_sn_bao_twoflat} {\it
Top}: Comoving distances of two flat $\Lambda$CDM models fitted with
the SN and BAO data. {\it Bottom}: Percent-level discrepancy in
comoving distances for the two models.}

\centering
\includegraphics[width=0.7\textwidth, angle=0]{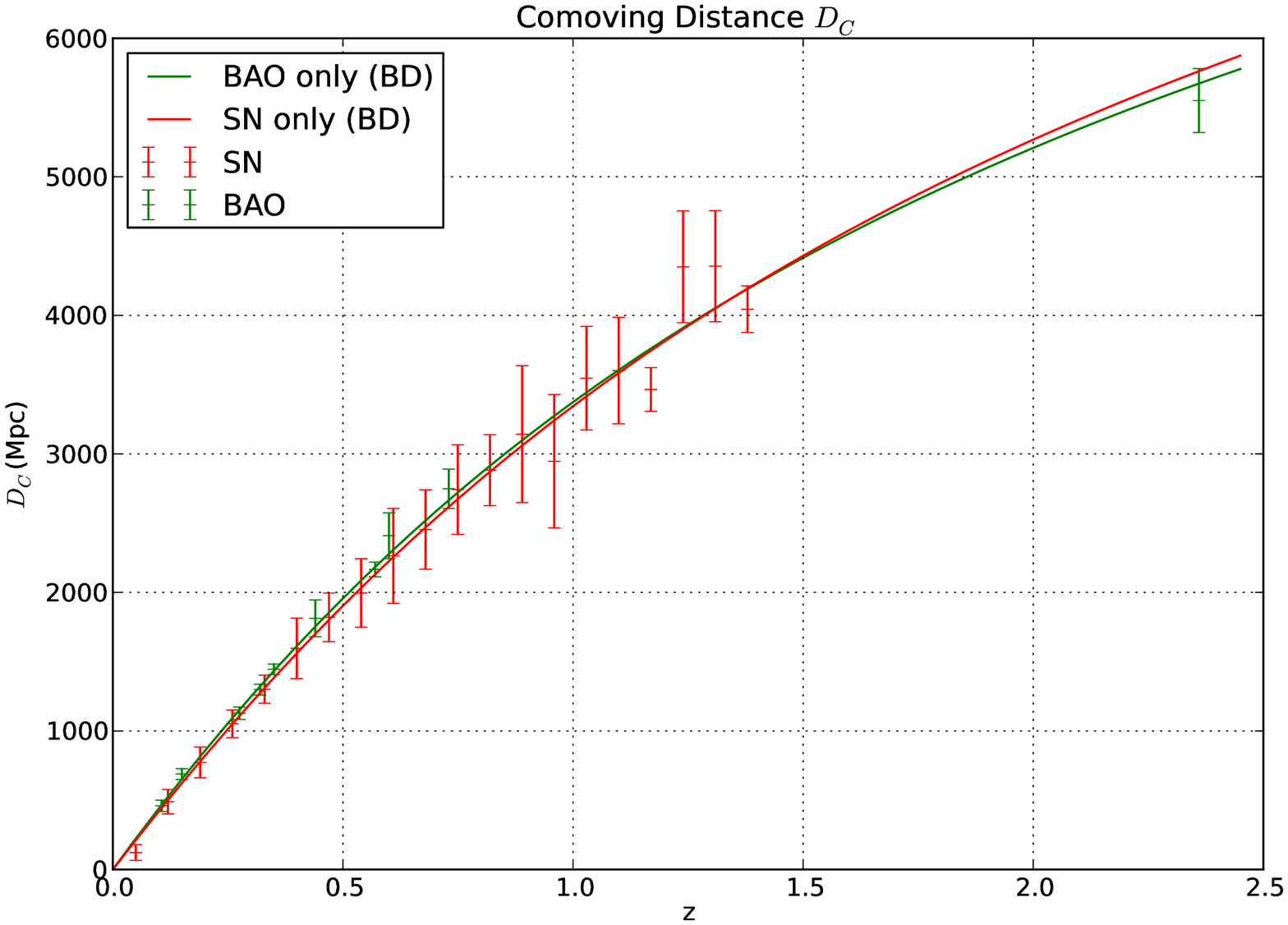}

\vspace{-3mm}

\caption{\baselineskip 3.6mm \label{fig:sn_bao_separate} Fitting the
BD model with SN{Ie} data and BAO data separately. Red dots with
error bar{s} are the SN data grouped into 20 bins. Green dots are
BAO data. {The r}ed curve is the fitted curve of SNe{Ia}. {The
g}reen curve is the fitted curve of BAO. {The best fit parameters
for }SN{e}{Ia}{ are} $(\Omega_\Lambda = 0.7326, h = 0.7174, \gamma_G
= 7.1143, \zeta = 0.0027)${. The best fit parameters for} BAO{ are}
$(\Omega_\Lambda = 0.6398, h = 0.6623, \zeta = 0.0096)${.} }

\end{figure}
\begin{figure}
\centering
\includegraphics[width=0.7\textwidth, angle=0]{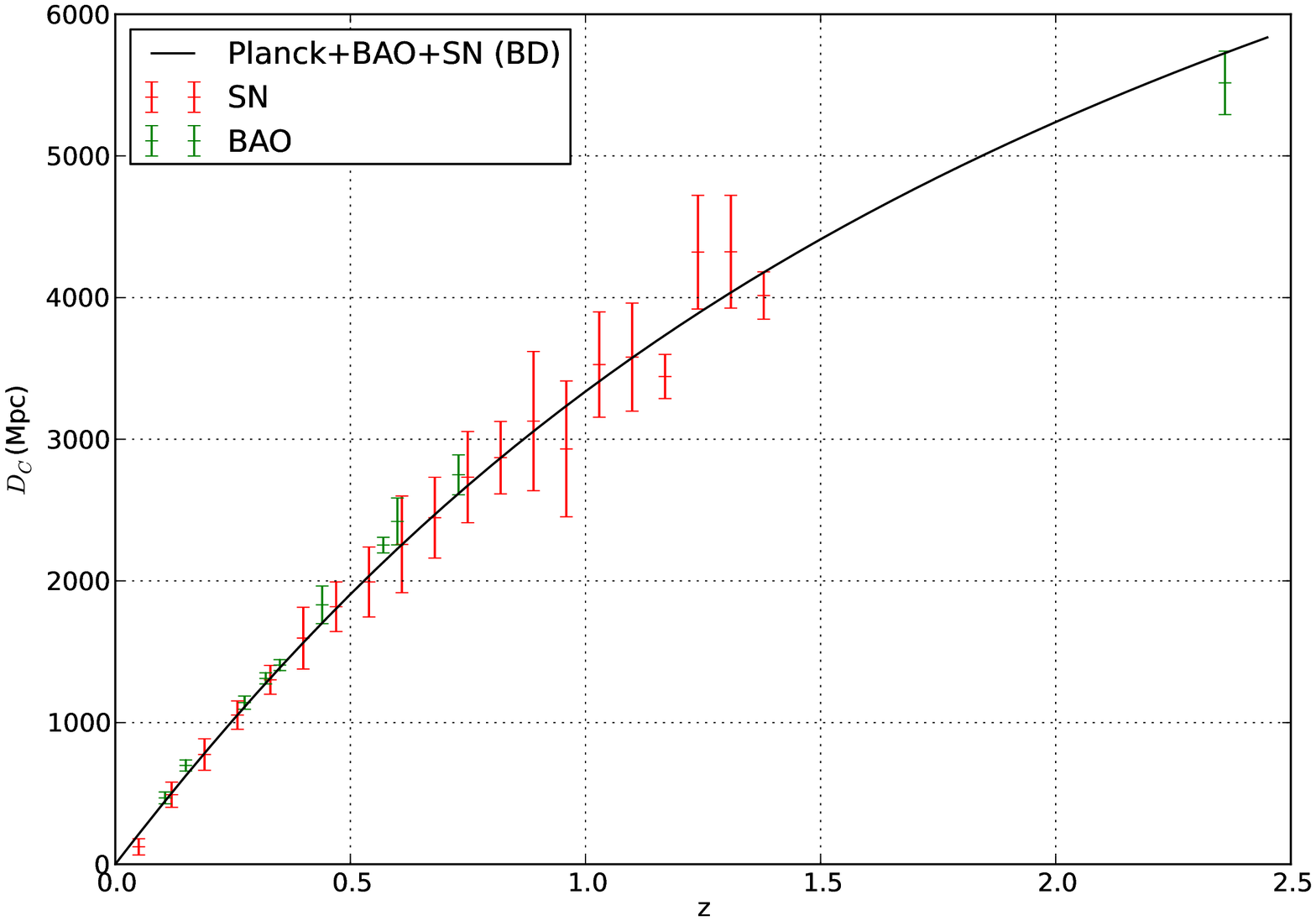}\\

\vspace{-3mm}

\caption{\baselineskip 3.6mm \label{fig:dC_joint_constrain} The
comoving distances of SN and BAO with a joint constraint of
(Planck+BAO+SN{Ia}){. T}he data point{s} representing $D_C$ given in
this plot {have been} corrected with the best-fit model. }
\end{figure}

In Figure~\ref{fig:dC_sn_bao_twoflat} we plot the fitting to the GR
($\Lambda$CDM) model with the current SNIa data and BAO data sets.
This curve show{s} that there is a small (percent-level)
difference between the best fit of the two data sets, with the BAO
data giving slightly larger distances. While it is quite probable
that this difference is simply due to statistical fluctuation, the
luminosity difference{ induced by varying $G$} in the SNIa also provides
a possible alternative explanation.

In Figure~\ref{fig:sn_bao_separate}, we plot the redshift-comoving
distance relation for the best-fit BD model with the BAO data and
the (binned) SNIa data separately. We see that although in this case
two different data sets are used, the best-fit models give nearly
identical $D_C(z)$ curves. This is what we would have expected,
because in the SN fit the {distance} measured {with BAO} is used
implicitly, and $\gamma_G$ is varied as a free parameter to
accommodate the differences between the two.

We then make a joint constraint on the BD model with the
Planck+BAO+SNIa data sets, allowing a free $\gamma_G$. The result is
shown in Figure~\ref{fig:dC_joint_constrain}. Here, the SNIa $D^C_L$
is corrected so that {these values} can be compared with the BAO
data. The redshift distance curve for the joint best-fit model is
plotted. We see again that the correction is not large in the end.

These results show that if variation in SNIa luminosity is
considered and treated as {a }free parameter, small differences
between the distances measured with{ separate} BAO and SNIa data
set{s} {can} be reconciled.{ }At present, however, this difference
is fairly small, and GR works fine,{ so} one can then constrain the
BD model instead.

\subsection{Model and Parameter Constraints}

In the above, we showed the redshift-distance relation for the
best-fit models with various given conditions, in order to
illustrate our discussion on effect of SNIa luminosity evolution due
to {the }change of $G$. Here we show how these global fittings were
done, and constraints on the BD parameter and other cosmological
parameters were obtained with cosmological observational data. To
derive such constraints on the BD model, we use the publicly
available \texttt{COSMOMC}
 code \citep{2002PhRvD..66j3511L} which implement{s} {a} Markov-Chain Monte Carlo (MCMC) simulation to
explore the parameter space and obtain limits on cosmological
parameters. The code {was} modified earlier by
\citet{2010PhRvD..82h3003W} to calculate the cosmic evolution with
BD gravity. In the current work, we updated the \texttt{COSMOMC}
code with newer versions, and also include the new data. Fitting
the Planck, BAO and SNIa datasets, we obtain the best fitting
results{ that are} summarized in Table~\ref{tb:fit_params}.

\begin{figure}
  \begin{center}
\includegraphics[width=0.45\textwidth]{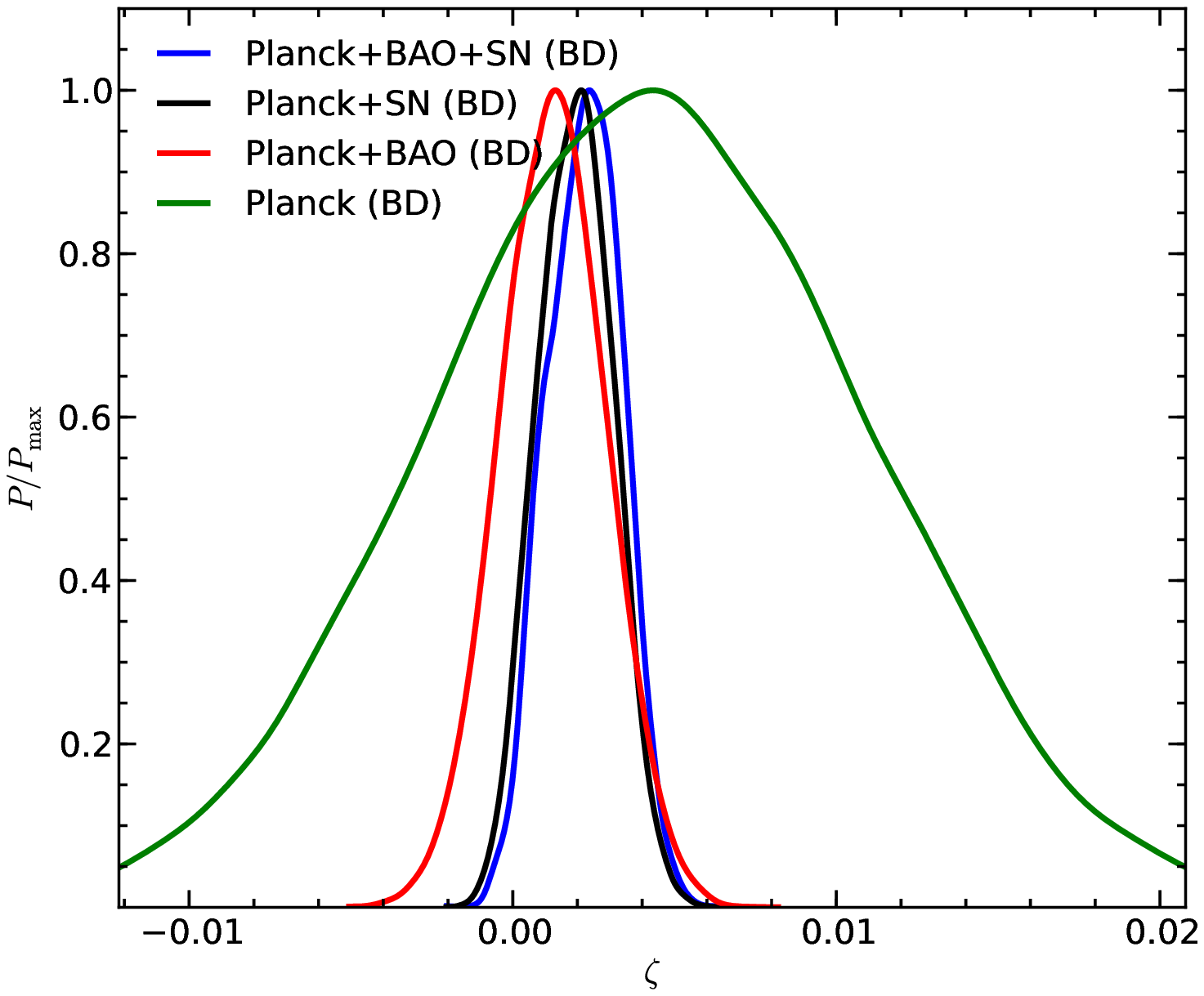}

\caption{ \baselineskip 3.6mm \label{fig:zeta_1d}
 The one dimensional likelihood distribution for $\zeta$.
``Planck'' denotes the result of only using Planck temperature data.
 ``Planck+BAO'' denotes the combined constraint with
 BAO data mentioned in Sect.~\ref{sec:BAO_data}.
``Planck+BAO+SN'' represents the result after adding{ the} updated
Union2.1{ SNIa data} described in Sect.~\ref{sec:SNIa_data}. }
\end{center}
  \begin{center}
\includegraphics[width=6.5cm, angle=0]{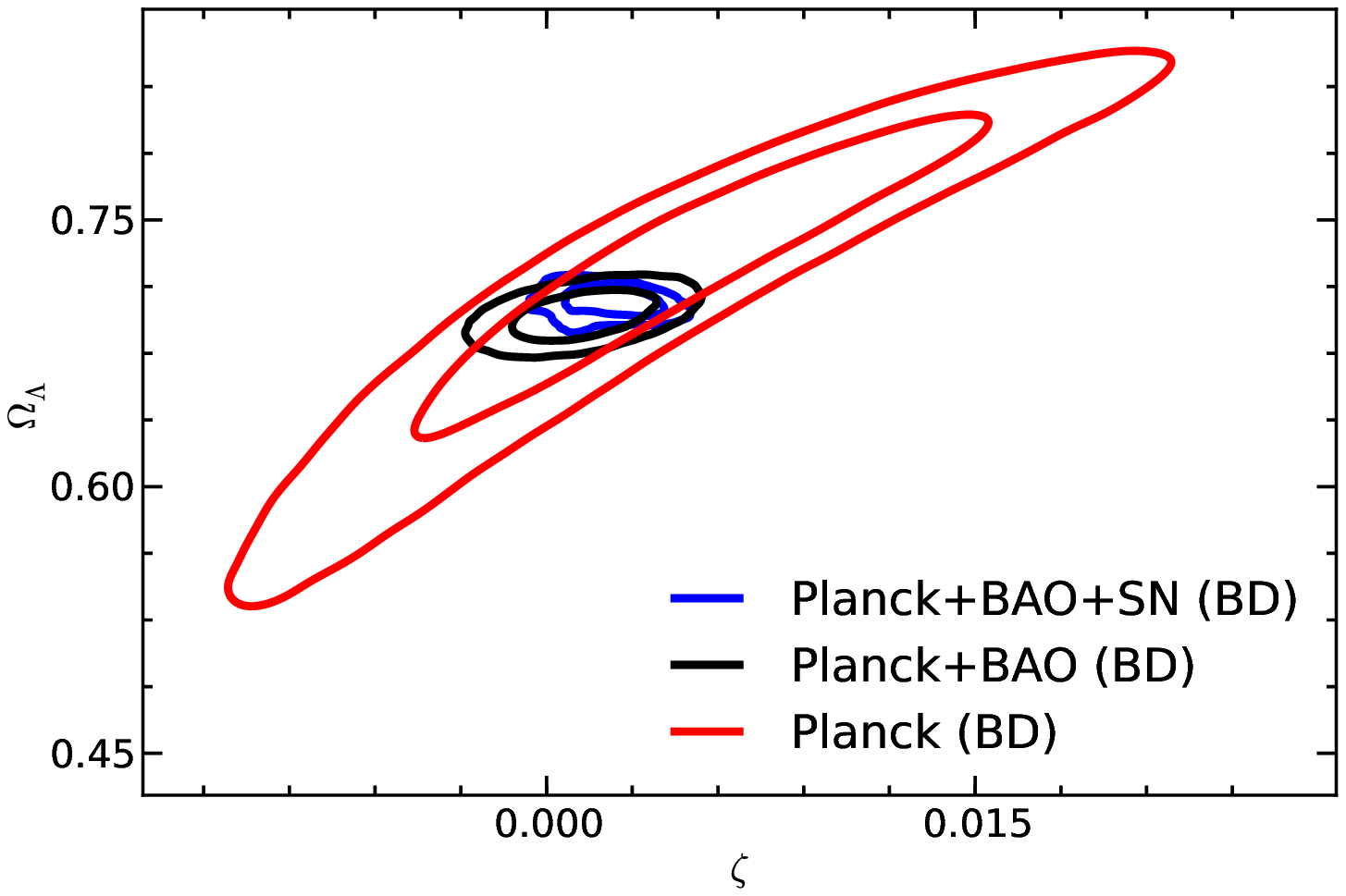}
\includegraphics[width=6.5cm, angle=0]{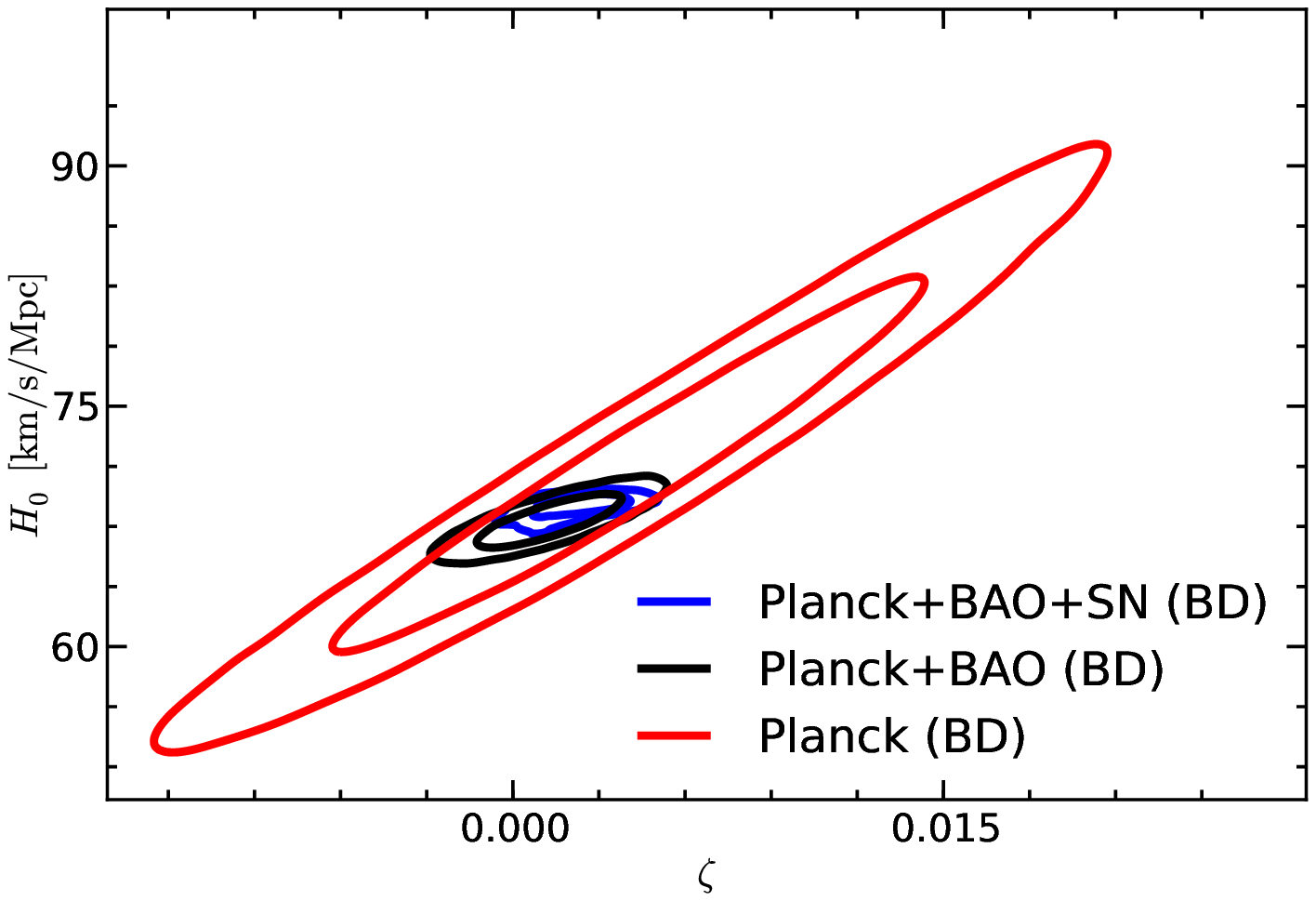}

\vspace{-3mm} \caption{\baselineskip 3.6mm \label{fig:zeta_2d}{\it
Left}: The two dimensional contour{s} for $\zeta$ {versus}
$\Omega_{\Lambda}$. {\it Right}:{ }The two dimensional contour for
$\zeta$ {versus} $H_0$. }
\end{center}
\end{figure}

In Figure~\ref{fig:zeta_1d}, we plot the 1-D marginalized
distribution for the re-parameterized BD model parameter $\zeta$
with different combination{s} of data sets. The curve which is
labeled ``Planck'' is the result{ of only} using the Planck CMB
temperature data. This distribution is relatively broad, as
degeneracy of parameters limited the precision of this test. The
curve which is labeled ``Planck+BAO'' {plots a combination of} the
CMB data {and} the BAO observation{al} data mentioned in Section
\ref{sec:BAO_data}, which yield{s} a much tighter constraint.  The
``Planck+BAO+SN'' shows the result combining the Planck, BAO and
updated Union2.1{ SNIa} data together, which is even tighter than
the Planck+BAO case, though not by much. Interestingly, the peak of
the distribution deviates from the one for {the }Planck+BAO case,
indicating that the {SNe} could significantly change the result. In
the plot, the peaks of the probability distributions are all at
$\zeta>0$, {slightly }favoring the model with $\omega>0$, and
especially so when the SNIa data are included in the fit. However,
the GR case ($\zeta=0$) {is} still within the limit, so the
fitting{s} are consistent with GR.

Figure~\ref{fig:zeta_2d} shows the two dimensional contours for
$\zeta$ {versus} $\Omega_{\Lambda}$ and $H_0$. In both cases, we can
see that if only the CMB data from Planck {are} used, there is
significant degeneracy as the contours extend to a long
{``}banana" shape. With the addition of the BAO and SNIa
data, the degeneracy{ in the parameters} is broken, resulting in much stronger
constraints on these parameters.

For the{ case of the} three{ combined} data set{s} (Planck+BAO+SN), we find
the 68\% and 95\% intervals are
\begin{eqnarray}
0.08\times10^{-2} &< \zeta <& 0.33\times10^{-2} ~~(68\%)\ ; \label{eq:zeta_3data}\\
-0.01\times10^{-2} &<\zeta <& 0.43\times10^{-2} ~~(95\%)\ .
\end{eqnarray}
These correspond to
\begin{eqnarray}
   1249.50 >  \omega > 302.53 ~~(68\%)\ ; \\
(\omega<-9999.50)  \cup  (\omega > 232.06) ~~(95\%)\ . \label{eq:zeta_3data_1}
\end{eqnarray}
Comparing this with what we {obtained} using the
Planck+BAO data,
\begin{eqnarray}
-0.04\times10^{-2} &< \zeta <& 0.28\times10^{-2} ~~(68\%)\ ; \\
-0.19\times10^{-2} &<\zeta <& 0.44\times10^{-2} ~~(95\%)\ {,}
\end{eqnarray}
which correspond to
\begin{eqnarray}
(\omega<-2499.50) &~\cup~& (\omega > 356.64) ~~(68\%)\ ; \\
(\omega<-525.82) &~\cup~& (\omega > 226.77) ~~(95\%)\ ,
\label{eq:zeta_bao}
\end{eqnarray}
the result is slightly{ improved}.

We can also derive limits on the variation of the gravitational
constant using {this} data set. Note that such limit{s} {are}
somewhat model-dependent, nevertheless {they} could give an idea
{about} current precision. To do this, we outputted two derived
parameters from the MCMC code, i.e. $\dot{G}/G \equiv
-\dot{\varphi}/\varphi$, which is the rate{ of change} of{ the}
gravitational constant at present, and $\delta G/G \equiv (G_{rec} -
G_0)/G_0 $, which is the integrated change {in the} gravitational
constant since the epoch of recombination. For the Planck+BAO+SN
case, we obtain
$$\dot{G}/G = -0.2649\times10^{-12}, \qquad \delta{G}/G =  0.0189 $$
and the $68\%$ marginalized limits are
\begin{eqnarray}
-0.3616\times10^{-12} < &\dot{G}/G& < -0.0820 \times10^{-12}  \\
0.0060 < &\delta G/G& < 0.0258\,. \label{eq:GCMB+BAO}
\end{eqnarray}

\begin{table}[h!!]
  \centering

  \begin{minipage}{105mm}
    \caption{ \label{tab:G} Constraints on $\dot{G}/G$.
The Errors are $1\sigma$ Unless Otherwise Noted }\end{minipage}

\renewcommand\baselinestretch{1.12}
    \fns\tabcolsep 4mm
    \begin{tabular}{ll}   \hline\noalign{\smallskip}
      $\dot{G}/G~[10^{-13} $yr$^{-1}]$   & Method \\
       \noalign{\smallskip}\hline\noalign{\smallskip}
       $2\pm 7$ &lunar laser ranging \citep{Muller:2007zzb}\\
       \noalign{\smallskip} \hline\noalign{\smallskip}
        $0\pm 4$ &{B}ig {B}ang nucleosynthesis (\citealt{Copi:2003xd}; \citealt{Bambi:2005fi}) \\
        \noalign{\smallskip} \hline\noalign{\smallskip}
         $0\pm 16$  & helioseismology \citep{Guenther98} \\
          \noalign{\smallskip}\hline\noalign{\smallskip}
          $-6\pm20$&neutron star mass \citep{Thorsett:1996fr} \\
           \noalign{\smallskip}\hline\noalign{\smallskip}
           $20\pm40$ &Viking lander ranging \citep{hellings1983} \\
            \noalign{\smallskip}\hline\noalign{\smallskip}
            $40 \pm 50$ & binary pulsar \citep{Kaspi:1994hp} \\
             \noalign{\smallskip}\hline\noalign{\smallskip}
             $-96\sim 81 ~(2\sigma)$& CMB (WMAP3) \citep{Chan:2007fe} \\
              \noalign{\smallskip}\hline\noalign{\smallskip}
              $-17.5\sim 10.5 ~(2\sigma)$& WMAP5+SDSS LRG  \citep{2010PhRvD..82h3003W} \\
               \noalign{\smallskip}\hline\noalign{\smallskip}
               $-1.42^{+2.48}_{-2.27} ~(1\sigma) $& Planck+WP+BAO \citep{2013PhRvD..88h4053L} \\
               \noalign{\smallskip}\hline\noalign{\smallskip}
               $-2.65^{+1.83}_{-0.97} ~(1\sigma) $& Planck+BAO+SN (This paper) \\
               \noalign{\smallskip}\hline
                            \end{tabular}

                 \end{table}

An updated summary of the various constraints on $\dot{G}/G$ with
different methods is given in Table~\ref{tab:G}. 
 We
see that compared with {the }other method, including high
precision solar system experiments, the cosmological constraint we
obtain is quite competitive.

\begin{figure}[tp]
  \begin{center}
\includegraphics[width=13.8cm, angle=0]{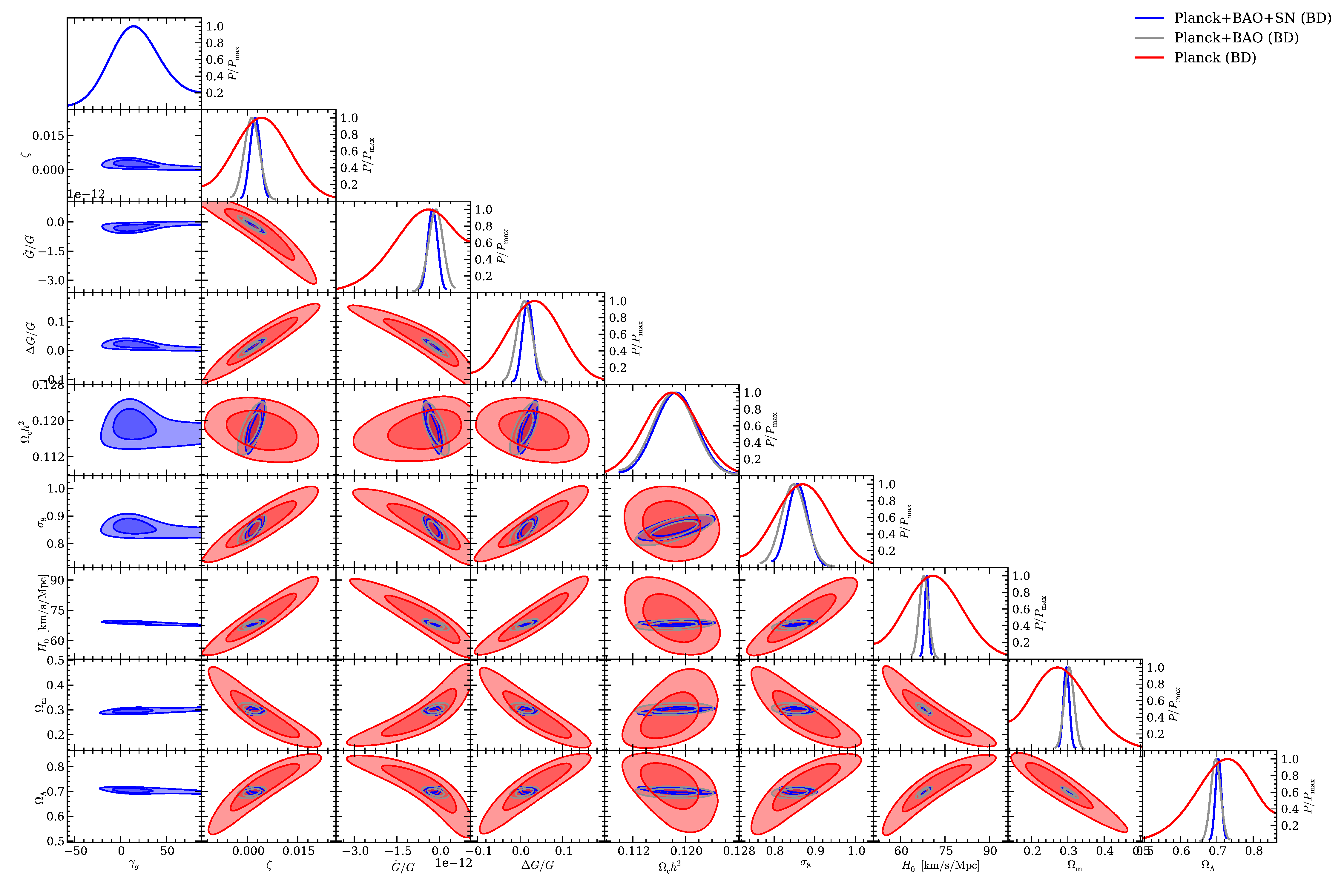}

\caption{\baselineskip 3.6mm \label{fig:triangle}The contours of{
the} marginalized distribution for the parameters in the BD theory.
}
\end{center}
\end{figure}

\begin{table}[htp]
\centering

\begin{minipage}{125mm}

\caption{\baselineskip 3.6mm \label{tb:fit_params} Fit and Limits
on Cosmological Model Parameters{ Obtained} with{ the} BD Model}
\end{minipage}


\renewcommand\baselinestretch{1.6}
\scriptsize \tabcolsep 1.4mm
\begin{tabular}{l|cc|cc|cc} \hline
& \multicolumn{2}{|c}{Planck+BAO+SN (BD)}&
\multicolumn{2}{|c}{Planck+BAO (BD)}& \multicolumn{2}{|c}{Planck (BD)}\\
\hline
Parameter & Best fit & $68\%$,~$95\%$ limits& Best fit & $68\%$,~$95\%$ limits& Best fit & $68\%$,~$95\%$ limits\\
\hline
$\gamma_G$  & $ 67.7232$ & $ 10.5205_{ -8.6986-73.7611}^{+ 42.6913+ 73.2123}$ &  & &  & \\
$\zeta$  & $  0.0005$ & $  0.0024_{ -0.0016 -0.0025}^{+  0.0009+  0.0019}$ & $  0.0017$ & $  0.0013_{ -0.0017 -0.0032}^{+  0.0015+  0.0031}$ & $  0.0040$ & $  0.0043_{ -0.0068 -0.0129}^{+  0.0065+  0.0128}$ \\
$\Omega_{\mathrm{c}} h^2$  & $  0.1172$ & $  0.1183_{ -0.0022 -0.0042}^{+  0.0022+  0.0044}$ & $  0.1184$ & $  0.1181_{ -0.0021 -0.0042}^{+  0.0024+  0.0046}$ & $  0.1187$ & $  0.1179_{ -0.0027 -0.0053}^{+  0.0027+  0.0054}$ \\
$\sigma_8$  & $  0.8474$ & $  0.8560_{ -0.0191 -0.0350}^{+  0.0177+  0.0353}$ & $  0.8605$ & $  0.8479_{ -0.0201 -0.0395}^{+  0.0216+  0.0422}$ & $  0.8648$ & $  0.8695_{ -0.0515 -0.1001}^{+  0.0535+  0.1048}$ \\
$\Omega_{\mathrm{b}} h^2$  & $  0.0216$ & $  0.0215_{ -0.0002 -0.0005}^{+  0.0003+  0.0005}$ & $  0.0217$ & $  0.0215_{ -0.0002 -0.0005}^{+  0.0002+  0.0005}$ & $  0.0215$ & $  0.0215_{ -0.0003 -0.0006}^{+  0.0003+  0.0006}$ \\
$H_0$ $^1$
  & $ 67.7563$ & $ 68.9783_{ -0.9628 -1.5877}^{+  0.2284+  0.6046}$ & $ 68.3716$ & $ 67.7557_{ -0.9612 -1.9476}^{+  1.1600+  2.2038}$ & $ 70.4907$ & $ 71.2350_{ -7.4578-13.8141}^{+  7.5226+ 15.3996}$ \\
$\Omega_{\mathrm{m}}$  & $  0.3023$ & $  0.2956_{ -0.0050 -0.0106}^{+  0.0073+  0.0140}$ & $  0.2997$ & $  0.3038_{ -0.0093 -0.0177}^{+  0.0089+  0.0181}$ & $  0.2821$ & $  0.2601_{ -0.0371 -0.0775}^{+  0.0856+  0.1695}$ \\
$\Omega_\Lambda$  & $  0.6977$ & $  0.7044_{ -0.0073 -0.0140}^{+  0.0050+  0.0106}$ & $  0.7003$ & $  0.6962_{ -0.0089 -0.0181}^{+  0.0093+  0.0177}$ & $  0.7179$ & $  0.7399_{ -0.0856 -0.1695}^{+  0.0371+  0.0775}$ \\
\hline
\end{tabular}
\parbox{110mm}{Notes: $^1$ $H_0$ is in units of
[km~s$^{-1}$~Mpc$^{-1}$].}
\end{table}

We also obtained{ the} best-fit and limits on various cosmological
parameters {by using} the BD model with combination{s of data sets
that included} Planck only, Planck+BAO and Planck+BAO+SN. These
results are given in Table~\ref{tb:fit_params}. 
 We
have already discussed the constraint{ provided by} the parameter
$\zeta$ or equivalently $\omega${.} {F}or other parameters{,} the
precision{s} of the constraints are generally comparable with the
GR case.  The contour plots for the model parameters are shown in
Figure~\ref{fig:triangle}. These plots show that with the addition
of the BAO and SNIa data, the constraints on the parameter space
could be greatly tightened.

In summary, we have update{d} the cosmological constraints {related
to} the BD theory with new observation{al} data including the Planck
observation of CMB, BAO observation by SDSS and WiggleZ, and also
the SNIa observation{s} using the Union2.1 sample. We added the SNIa
observation{s} to the data set used in this work, and also
considered how variation of $G$ may affect the SNIa peak luminosity.
We find the result is still consistent with {GR} within error
limits. We derived limits on $\zeta$ (or equivalently $\omega$) and
$\dot{G}$. For the combined fits,  the limit{s} {are} significantly
reduced.

\begin{acknowledgements}
Our MCMC computation was performed on the Laohu cluster {at National
Astronomical Observatories, Chinese Academy of Sciences}. This work
is supported by the Ministry of Science and Technology {of China
(}863 project{,} Grant {No. }2012AA121701{)}, the {National Natural
Science Foundation of China (}NSFC{,} Grant {Nos.} 11373030 {and
}11473044{)},{ and} by the Strategic Priority Research Program ``The
Emergence of Cosmological Structures'' of the Chinese Academy of
Sciences (Grant No. XDB09000000).
\end{acknowledgements}

\label{lastpage}

\end{document}